# Absorption-Ablation-Excitation Mechanisms of Laser-Cluster Interactions in a Nanoaerosol System

Yihua Ren[1], Shuiqing Li[1,*], Yiyang Zhang[2], Stephen D. Tse[3], Marshall B. Long[4]

[1] *Key Laboratory for Thermal Science and Power Engineering of Ministry of Education, Department of Thermal Engineering, Tsinghua University, Beijing 100084, China*

[2] *Institute of Nuclear and New Energy Technology, Tsinghua University, Beijing 100084, China*

[3] *Department of Mechanical and Aerospace Engineering, Rutgers, The State University of New Jersey, Piscataway, NJ 08854, USA*

[4] *Department of Mechanical Engineering & Materials Science, Yale University, New Haven, Connecticut 06520, USA*

\*Corresponding author.

Prof. Shuiqing LI. **Fax:** +86-10-62773384; **Email:** lishuiqing@tsinghua.edu.cn

**Abstract**

The absorption-ablation-excitation mechanism in laser-cluster interactions is investigated by measuring Rayleigh scattering of aerosol clusters along with atomic emission from phase-selective laser-induced breakdown spectroscopy (PS-LIBS). As the excitation laser intensity is increased beyond 0.16GW/cm$^2$, the scattering cross-section of TiO$_2$ clusters begins to decrease, concurrent with the onset of atomic emission of Ti, indicating a scattering-to-ablation transition and the formation of nanoplasmas. To better clarify the process, time-resolved measurements of scattering signals are examined for different excitation laser intensities. For increasing laser intensities, the cross-sections of clusters decrease during a single pulse, evincing the shorter ablation delay time and larger ratios of ablation clusters. Assessment of the electron energy distribution during the ablation process is conducted by non-dimensionalizing the Fokker-Planck equation, with analogous Strouhal $Sl_E$, Peclet $Pe_E$, and Damköhler $Da_E$ numbers defined to characterize the laser-induced aerothermochemical environment. For conditions of $Sl_E \gg 1$, $Pe_E \gg 1$, and $Da_E \ll 1$, the electrons are excited to the conduction band by two-photon absorption, then relax to bottom of the conduction band by collisional electron energy loss to the lattice, and finally serve as the energy transfer media between laser field and lattice. The relation between delay time and excitation intensity is well predicted by this simplified model with quasi-steady assumption.

Laser-cluster interactions are widespread across fundamental physical processes in many disciplines. Depending on the excitation laser intensity, such interaction can be used for characterization of particles/aggregates based on elastic and inelastic scattering ($10^2$~$10^8$W/cm$^2$)[1-4], determination of local chemical compositions based on laser-induced breakdown or aerosol-fragmentation spectroscopy ($10^8$~$10^{12}$W/cm$^2$)[5-8], and investigation of laser-driven nonlinear clusters dynamics based on generation of energetic photons and X-rays ($10^{11}$~$10^{18}$W/cm$^2$)[9-15]. For scattering-to-breakdown transition, the generally-accepted mechanism involves production of initial electrons from multi-photon excitation or tunnel ionization (distinguished by the Keldysh adiabaticity parameter[10,16]), followed by fast production of electrons due to cascade collision ionization (inverse Bremsstrahlung) and emergence of shock wave(s) by hydrodynamic or Coulombic expansion. In practice, the breakdown threshold is defined by the excitation laser intensity at which the transmitted laser intensity decreases and the emission of Bremsstrahlung radiation forms. Recently, a new phase-selective laser-induced breakdown spectroscopy (PS-LIBS) has been developed for the diagnosis of gas-to-particle transition at nanoscale[17,18]. The PS-LIBS only excites constituent atoms (e.g. Ti in $TiO_2$) in the particle phase, with no breakdown emission occurring for surrounding gas molecules, presenting a robust technique for cluster/nanoparticle identification, monitoring, and concentration mapping for many aerosol systems. Spatially localized nanoplasmas are found around individual $TiO_2$ nanoparticles, without macroscopic sparks or Bremsstrahlung radiation, while atomic emissions are detected, implying a novel laser-cluster interaction mechanism between the scattering and breakdown regimes.

Such localized nanoplasmas formed around nanoparticles in PS-LIBS are believed to be produced through thermal ablation of the clusters. The thermal-ablation-driven laser-cluster interaction differs significantly from the laser-induced damage of solid materials and micro-sized particles because (*i*) impact ionization (inverse bremsstrahlung heating) is negligible due to the rare diffusion of electrons to higher energy levels[19] and (*ii*) avalanching explosive vaporization on the surface of micro-particles has not been observed[20]. Moreover, the novel interaction mechanism is different from those cases where intense laser photons strike Van der Waals crystal clusters given the quasi-steady feature of the thermal ablation process and the lack of hydrodynamic or Coulombic expansion due to the small multi-photon ionization rates of electrons[10]. Similar ablation phenomena have also been observed in laser-induced incandescence (LII) of soot[21] and metal oxides clusters[22,23], although these studies focused on the removal of material and the influence on particle irradiation. Li *et al.*[24] and Lucas *et al.*[8] proposed that lattice defects or surface excitons facilitate electronic excitation with photons of sub-band-gap energy in their investigations on the photo-fragmentation of wide-band-gap particles by UV light. However, as for the weak thermal ablation of narrow-band-gap semiconductor clusters without shock wave ahead of ejected species, a clearer physical picture is needed to better understand this laser-cluster interaction regime.

Here, the absorption-ablation-excitation mechanism in PS-LIBS is investigated by examining Rayleigh scattering and atomic emissions from clusters, with further analysis of a dimensionless Fokker-Planck equation. The physical mechanism involved in PS-LIBS is illustrated in Fig.1, along with the experimental setup. The laser intensity at the scattering-to-ablation transition point is clearly identified by the reduction of scattering cross-sections of clusters and the onset of atomic emission. The ablation delay time can be deduced by time-resolved scattering measurements, and can be modeled by appropriately non-dimensionalizing the Fokker-Planck Equation.

The experiment setup is similar with that employed in our recent work[18], with the schematic shown in Fig.1(d) and more details given as supplementary materials. The 532nm laser beam focuses on the centerline of the cluster-laden flow at 21mm above the burner exit. The signal is then collected into a spectrometer (Acton SpectraPro 300i) and detected by a PI-MAX3 ICCD camera. The ICCD gate width is set as 200ns for measuring the integrated Rayleigh scattering signal and then at 2.54ns (the minimum gate width) for measuring the temporal evolution of the scattering. By shifting the 2.54ns ICCD gate to different delay times after the laser pulse, the time evolution of the different collected spectra can be quantified. The timing of the ICCD gate and the excitation laser pulse during time-resolved measurement is monitored by a photodiode connected to an oscilloscope. The laser energies used here range from 0.1 to 120mJ/pulse, corresponding to average laser intensities from 0.02~24.4GW/cm2 at the focal point based on a Gaussian-distributed 1/e waist diameter of ~250μm. A series of neutral-density filters are placed before the focal lens to adjust the excitation laser intensity with little change to the profile and delay time of the laser pulse. The flame-synthesized clusters have an average diameter of ~11nm, with a number density of ~$10^{11}$/cm$^3$, as modeled by population balance and further confirmed by in-situ TEM sampling. Due to the large number density of clusters, the measurement of responses of clusters upon repetitive laser pulse excitation reach statistical significance.

Scattering intensity, scattering efficiency, and atomic emission of clusters for varying excitation laser intensity are shown in Fig.2. The exposure time is 20s (corresponding to 200 laser shots). The scattering response from pure clusters can be calculated by subtracting scattering signals of nano-aerosols (gas+cluster) from that of pure gases (gas). The flame environment does not change with or without clusters ensuring the same gas Rayleigh cross-sections for the two situations. The scattering intensity of clusters increases proportionally with laser intensity, and then flattens out, while the scattering intensity of gases is linear with laser intensity, for the same range, as depicted in the inset of Fig.2. The scattering efficiency of clusters, defined as the ratio of the scattering signal intensity over the laser intensity, is approximately constant up to 0.8mJ/pulse (1.6GW/cm$^2$), and then begins to decrease after this critical value, indicating the same tendency of scattering cross section and demonstrating the onset of laser ablation of the clusters. Above the ablation threshold for these conditions, the atomic emission from Ti atoms, *i.e.* the PS-LIBS, is observed, further corroborating the formation of nanoplasmas upon ablation. The atomic spectrum of Ti near 500nm is shown in the plot of Fig.1(c). It should be noted that the first appearance of atomic emission of Ti at the wavelength of 498.17nm (corresponding to the transition of electronic energy level $3d^34p$ to $3d^34s$) occurs exactly at the same laser intensity when clusters' cross-sections start to decrease, indicating that the PS-LIBS signal is caused by the ablation of clusters. The atomic emission intensities saturate after 1GW/cm$^2$, implying that the number of electrons after ablation plateaus at strong laser intensity, which will be discussed later.

The ablation process of clusters is further examined by time-resolved measurements, as shown in Fig.3. The solid red curves are the Gaussian fittings of the Rayleigh scattering signals from gases as a function of delay time, while the symbols represent the scattering intensities from clusters. The shot-to-shot fluctuation of the laser is less than 25% for intensity and less than 1.5ns for delay time, according to gas Rayleigh signals. After integration over 50 pulses, the laser temporal profile forms a perfectly Gaussian curve with errors less than 3% for intensity and 0.2ns for delay time. The small shot-to-shot fluctuations ensure the statistical significance of the accumulated time-resolved signal intensities from the probe volume. Different excitation laser intensities are investigated by adding neutral-density filters. Time-resolved scattering of both gases and clusters are normalized to unity at the maximum intensity. The relative time of the temporal evolution of the signals is set to zero, concurrent with an incoming laser pulse, *i.e.* the initial occurrence of the gas Rayleigh signal. When the laser intensities are below the ablation threshold, as seen in Fig.3(a) and Fig.3(b), the Rayleigh scattering of clusters is nearly synchronous with the gas Rayleigh signal, revealing the elastic response of the clusters to the laser pulse. When the laser intensities become larger than the ablation threshold, the scatterings/emissions of clusters deviate from that of gas, decreasing in the first few nanoseconds and then following the laser pulse after an obvious drop. The drop points of the cluster-scattering curves signify the reduction of cross-sections of clusters, *i.e.* the scattering-to-ablation transition of the clusters during the laser pulse. The atomic spectrum of Ti is also observed at the transition point. Therefore, the signal before the dropping point comes from the scattering response of clusters; while after the interruption point, the signal is then caused by the scattering of fragments after ablation and ensuring atomic emission. The elastic scattering of clusters before the transition point is fitted by Gaussian time-resolved distributions (as denoted by the red dashed curves in Fig.3), while the scattering of clusters after the transition point is quantified as the normalized gas Rayleigh curves. Consequently, the fragmentation degree during the ablation process can be characterized by the instantaneous ratio of the red-dashed-curve value to the red-solid-curve value. The fragmentation degree increases for larger laser intensities, which is consistent with the decrease of integrated cross-sections of clusters at larger laser intensities as shown in Fig.2. The ablation delay time, which is marked in Fig.3 as the obvious dropping point of cluster scattering, becomes shorter with increasing laser intensities for the range examined.

According to semiconductor absorption theory[25], the direct absorption of 2.34eV photon by $TiO_2$ nanoparticles is difficult because *(i)* the photon energy is below the bandgap~3.2eV, *(ii)* the lattice absorption region locates at the infrared region and *(iii)* the surface effect does not enhance the absorption of 532nm light significantly[26]. Therefore, it is believed that conduction-band electrons from multi-photon excitation are responsible for the ablation process. The energy distribution of electrons can be described by the Fokker-Planck Equation[19,27], *i.e.*, the convection-diffusion equation in energy space:

$$\frac{\partial}{\partial t}f(u,t) = \frac{\partial}{\partial u}\left[(B_{u,t} - A_{u,t})f(u,t) + D_{u,t}\frac{\partial f(u,t)}{\partial u}\right] + S, \qquad (1)$$

where $f(u,t)du$ is the number of electrons with energy between $u$ and $u+du$ at time $t$, and $S=S_{MPI}+S_{imp}$ represents the sources and sinks of electrons considering multi-photon excitation and impact ionization (the recombination term can be neglected given the relatively large band-gap of 3.2eV for $TiO_2$ considered here). The first term on the right is the net number of electrons across an energy value $u$ per unit time, including the convection and diffusion of conduction-band electrons in energy space. The convection term contains the rate of absorption of electromagnetic energy by electrons via collisions, *i.e.* joule heating rate, $A_{u,t}$, and the rate of electron energy loss to the lattice, $B_{u,t}$, for which the formulas are presented in detail in Refs.[19,27-29]. The Fokker-Planck Equation can be non-dimensionalized to the form:

$$\frac{\partial f^*}{\partial t^*}\frac{t_{conv}}{t_{laser}} = \frac{\partial}{\partial u^*}\left(f^* + \frac{t_{conv}}{t_{diff}}\frac{\partial f^*}{\partial u^*}\right) + \frac{t_{conv}}{t_{react} \cdot f_p}, \qquad (2)$$

where $t_{laser}$ is the time scale of the laser pulse(~10ns), $t_{conv}=E_{bg}/(B-A)$ is the convection time ($10^{-7}$~$10^{-6}$ns), $t_{diff}=E_{bg}^2/D$ is the diffusion time (0.1~$10^2$ns), $t_{react}=2h\nu/\beta I^2$ is the excitation time of two-photon absorption ($10^{-6}$~$10^2$ns) which is strongly dependent on the laser intensity from 0.02-20.4GW/cm$^2$ ($\beta$ is the two-photon absorption coefficient, $I$ is the laser intensity, $h\nu$ is the photon energy), and $f^*=f/f_p$ is the ratio of the number of conduction-band electrons over the number of molecules in one cluster. The Strouhal number $Sl_E$, defined as $t_{laser}/t_{conv}$ (analogous to that defined in fluid dynamics)[30], expresses the ratio of the intrinsic time scale to the convective time scale and is about $10^7$~$10^8$. Thus the whole ablation process reaches quasi-steady state. The Peclet number $Pe_E$, defined as $t_{diff}/t_{conv}$, is about $10^5$~$10^9$. Thus, diffusion of conduction-band electrons in energy space can be neglected, and impact ionization by diffused electrons to higher energies is not considered here. The dimensionless reaction parameter $Da_E$, defined as $t_{conv}/t_{react}$ (analogous to Damköhler number in combustion systems[31]), ranges from ~$10^{-5}$ at weak laser intensity (0.02GW/cm$^2$) to ~1 at strong laser intensities (20.4GW/cm$^2$). At moderate values of laser intensities, electrons relax to the bottom of the conduction band after two-photon excitation. Under $Sl_E \gg 1$, $Pe_E \gg 1$, $Da_E \ll 1$ conditions, the electrons are excited to the conduction band by two-photon absorption, and relax to bottom of the conduction band by electron energy loss via collisions with lattice, as previously described and depicted in Fig.1(a). This absorption-ablation-excitation laser-cluster interaction is different from laser-induced breakdown and intense laser-cluster interaction. For the laser-induced breakdown regime, computations from Holway[19] showed that $Pe_E$ reaches 1, and diffusion of electrons contributes to impact ionization. On the other hand, for the intense laser-cluster interaction regime, the assumption of $Sl_E \gg 1$ is invalid due to the picosecond (or even femtosecond) laser pulse[10] and the whole process cannot be regarded as quasi-steady state. For the photofragmentation of metal particles, a large source of nearly free electrons leads to $Da_E \gg 1$.

Consequently, a simplified ablation model is proposed with the assumption that all the conduction-band electrons $N$ are created by two-photon excitation:

$$\frac{\partial N}{\partial t} = \frac{\beta I^2 V_p}{2h\nu}. \tag{3}$$

The cluster lattice is heated to vaporization by joule heating from conduction-band electrons under an electric field with the power of $AN$. The clusters are ablated in shorter time for stronger laser field due to the faster production of conduction-band electrons and the stronger joule heating, which is observed in the time-resolved measurement and well predicted by the model, as shown in Fig.4. The small deviation at large laser intensities may be caused by the less rigorous assumption of $Da_E \ll 1$ and the possibility that not all the electrons locate at the bottom of the conduction band. At the minimum ablation laser intensity, which is denoted by the vertical red dashed line, the delay time of ablation is about the duration of the laser pulse. Therefore, the scatter-to-ablation transition threshold near the laser intensity of 0.16 GW/cm² is mainly due to the ablation duration becoming longer than the laser pulse duration. The number of electrons increases and then saturates at large laser intensities shown in the blue dashed line in Fig.4. The modeled electron numbers during ablation can explain the trend of Ti atomic emission with increasing laser intensity to some extent. The electrons, once in the conduction band, are accelerated in the laser field and collide with the surrounding atoms and ions. Because the atomic emission signal is determined by both the number and energy level distribution of excited atoms, the saturation and even declining tendency of atomic emission in Fig.2 is partially caused by the saturation trend of electron numbers and partially by the further stepwise ionization of excited atoms in the nanoplasma at high laser intensities, which is similar to excitation and ionization in gaseous plasma[32]. This phenomenon needs further investigation.

In summary, the integrated and temporal Rayleigh scattering measurements, together with the model derived from the Fokker Planck equation, explore a new regime of laser-cluster interaction. Both the scattering signals and the atomic emissions point to the occurrence of ablation of clusters under the average laser intensity~0.16GW/cm$^2$. With time-resolved data and dimensionless analysis, the physical mechanism of the ablation process is clarified, where the electrons are first excited to the conduction band by two-photon absorption, then return to the bottom of the conduction band by collisional loss to the lattice, and finally become the energy transfer media between the laser field and the lattice. Once in the conduction band, the electrons are accelerated in the laser field and collide with the surrounding atoms and ions forming the basis for PS-LIBS.

This research is supported by the National Natural Science Funds of China (No.51176094) and by the National Key Basic Research and Development Program (No.2013CB228506). SQL acknowledges the China Scholarship Council Scholarship for his sabbaticals at Yale University and Princeton University, as well as Prof. Michael Renfro at UConn and Prof. Yikang Pu at Tsinghua for discussions.

# Figures:

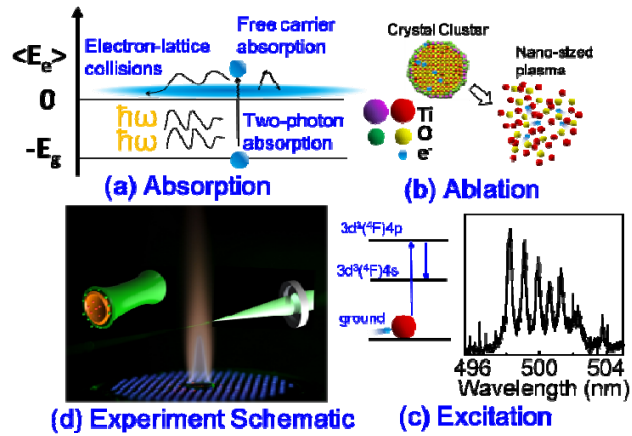

Fig.1. Schematic of (a-c) the absorption-ablation-excitation mechanism and (d) the experiment set-up. In panel(a), the electrons are excited to the conduction band by two-photon absorption, then return to the bottom of the conduction band by electron energy loss and transfer energy from laser to lattice. The clusters are ablated and then transform into nanoplasma, as depicted in panel(b). Finally, the electrons in the plasma excite the atoms producing atomic emission, as shown by the spectrum in panel(c).

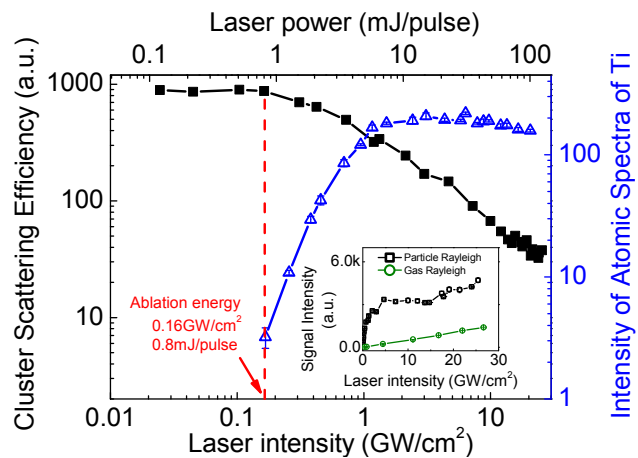

Fig.2. Scattering efficiency of $TiO_2$ clusters (black squares) and atomic emission intensity of Titanium at 498.17nm(blue triangles), as a function of excitation laser intensity. The scattering efficiency of clusters is defined as the ratio of the scattering signal intensity over the laser intensity. The inset shows the comparison of scattering signals of clusters (black diamonds) and gases (green circles).

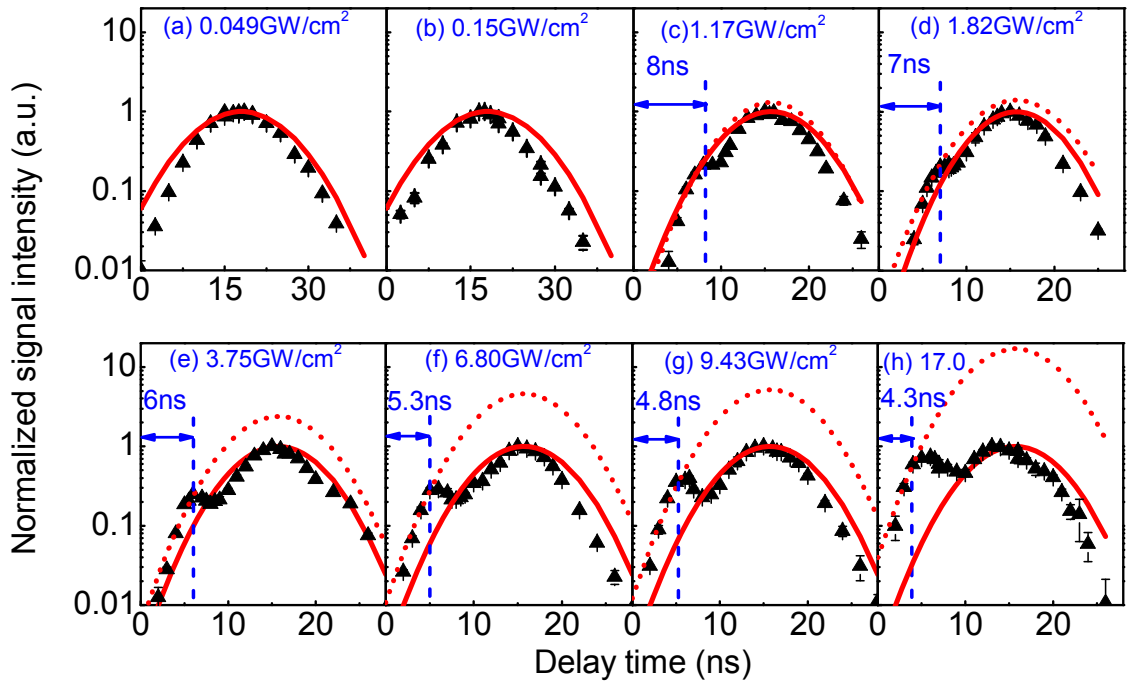

Fig.3. Time-resolved measurement of scattering signals of clusters (black triangles) and Gaussian fitting of Rayleigh scattering (red curves) at different laser intensities. The elastic scatterings of clusters before the transition point are fitted by Gaussian curves (dashed curve), which extrapolate beyond the transition point, with similar profiles but different amplitudes compared to the gas Rayleigh curves.

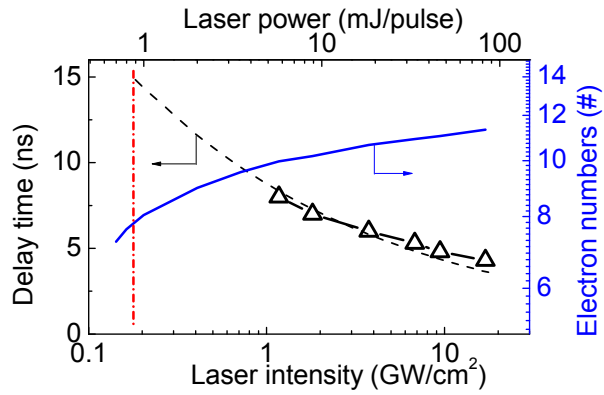

Fig. 4. The ablation durations at different laser intensities are denoted by the black line (model result) and the symbols (time-resolved experimental measurements). At the minimum ablation laser intensity, which is marked by the red dashed line, the delay time of ablation is about same as the duration of the laser pulse. The modeled electron number is characterized by the blue dashed line, which increases and saturates with increasing laser intensity.